\documentclass[aps,twocolumn,showpacs]{revtex4}
\usepackage{amsmath}
\usepackage{graphicx}

\begin{document}

\title{High Energy Particles in the Solar Corona}

\author{A. Widom}
\affiliation{Physics Department, Northeastern University, Boston MA USA}
\author{Y.N. Srivastava}
\affiliation{Physics Department \& INFN, University of Perugia, Perugia IT}
\author{L. Larsen}
\affiliation{Lattice Energy LLC, 175 North Harbor Drive, Chicago IL USA}

\begin{abstract}
Collective Ampere law interactions producing magnetic flux tubes   
piercing through sunspots into and then out of the solar corona allow for 
low energy nuclear reactions in a steady state and high energy 
particle reactions if a magnetic flux tube explodes in a violent event 
such as a solar flare. Filamentous flux tubes themselves are 
vortices of Ampere currents circulating around in a tornado fashion  
in a roughly cylindrical geometry. The magnetic field lines are parallel 
to and largely confined within the core of the vortex. The vortices may 
thereby be viewed as long current carrying coils surrounding magnetic flux 
and subject to inductive Faraday and Ampere laws. These laws set the energy 
scales of (i) low energy solar nuclear reactions which may regularly occur and 
(ii) high energy electro-weak interactions which occur when magnetic flux coils 
explode into violent episodic events such as solar flares or coronal mass 
ejections.    
\end{abstract}
\pacs{94.20.wq, 96.25.Qr, 96.60.P-, 96.60.Hv}
\maketitle

\section{\label{intro} Introduction}

\begin{figure}[tp]
\scalebox {0.6}{\includegraphics{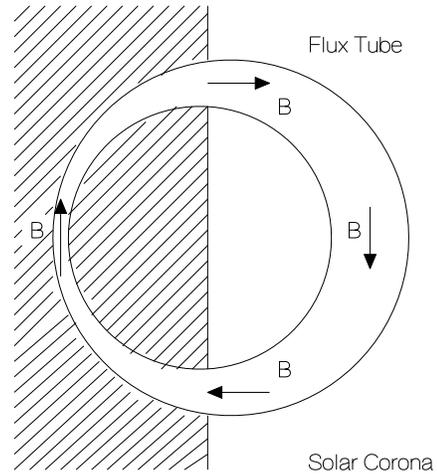}}
\caption{Shown schematically is a magnetic flux tube which exits the solar 
photosphere (shaded region) and enters the solar corona (clear region) through 
a sunspot on the chromosphere boundary. The magnetic flux tube then exits the 
solar corona and enters back into the photosphere through a second sunspot on 
the chromosphere boundary. The walls of magnetic flux tube are a vortex 
of circulating electric currents.}
\label{fig1}
\end{figure}

For physical reasons which are presently not entirely clear, dark sunspots exist 
on the optical solar surface. These have long been observed ever since Galileo saw 
sunspots with his optical telescope. It was later found that magnetic flux tubes 
exit out of some solar sunspots and enter back into 
others\cite{Hale:1908,Hale:1919}. Employing modern X-ray telescopes,
spectacular pictures have been taken of magnetic flux 
tubes\cite{Priest:1998} which arch into and out of the solar corona,
well above the optical solar surface. The situation is schematically shown in 
FIG.\ref{fig1}. The closed magnetic flux tubes are pictured in both the solar 
photosphere and the solar corona. The closed flux tube magnetic field lines enter 
into the solar corona through one sunspot and exit out of the solar corona 
through another sunspot. The floating flux tubes in the solar corona are held up by 
a magnetic buoyancy\cite{Parker:1955}. The outer walls of the magnetic flux tube consist 
of large circulating electric currents forming a turbulent vortex with a darker 
comparatively quiet magnetic core. When the magnetic flux tubes explode\cite{Parker:1957} 
into a solar flare, with or without a coronal mass ejection\cite{Kahler:2007}, 
charged particles with very high energy are  produced\cite{Forbush:1946,Dorman:1993,Reams:2004,Belov:2005,Vashenyuka:2005,Bostanjyan:2007}, 
say up to \begin{math} \sim 10^2 \end{math} GeV.  These relativistic particles can escape 
the sun and be observed on earth as ground level cosmic ray 
enhancements\cite{Shea:2001,Cliver:2006,L3:2006} induced by solar flares.

Our purpose is to discuss how such energetic particles arise in the solar corona 
and how these particles induce nuclear reactions well above the solar 
photosphere. The central feature of our explanation, which centers around 
Faraday's law, is the notion of a solar accelerator closely analogous to the 
betatron\cite{Kirst:1941,Serber:1941}. Conceptually, the betatron is a step up 
transformer whose secondary coil is a toroidal ring of accelerating charged 
particles circulating about a Faraday law (time varying) magnetic flux tube.

Acting as a step up transformer, it is possible for a solar magnetic flux tube 
to transfer circulating charged particle kinetic energy upward from the 
photosphere to circulating charged particles located in the corona. Circulating 
currents located deep in the photosphere can be viewed conceptually as a net 
current \begin{math} I_{\cal P} \end{math} circulating around a primary coil.  
Circulating currents found high in the corona can be viewed as a net current 
\begin{math} I_{\cal S} \end{math} circulating around a secondary coil. 
If \begin{math} K_{\cal P} \end{math} and \begin{math} K_{\cal S} \end{math} 
represent, respectively, the charged particle kinetic energies in the primary and 
secondary coils, then one finds the step up transformer power equation 
\begin{math} 
\dot{K}_{\cal P}=V_{\cal P}I_{\cal P} =V_{\cal S}I_{\cal S}=\dot{K}_{\cal S} 
\end{math}, 
wherein \begin{math} V_{\cal P} \end{math} and \begin{math} V_{\cal S} \end{math} 
represent, respectively, the voltages across the primary and secondary coils. 
The total kinetic energy transfer  
\begin{math} 
\Delta K_{\cal P}=\int V_{\cal P}I_{\cal P}dt=\int V_{\cal S}I_{\cal S}dt=\Delta K_{\cal S} 
\end{math}. 
The essence of the step up transformer mechanism is that the kinetic energy distributed 
among a very large number of charged particles in the photosphere can 
be transferred via the magnetic flux tube to a distributed kinetic energy shared among 
a distant much smaller number of charged particles located in the corona, i.e.
a small accelerating voltage in the primary coil produces a large accelerating voltage 
in the secondary coil. The resulting transfer of kinetic energy is 
{\em collective} from a large group of charged particles to a smaller group of charged 
particles. The kinetic energy per charged particle of the dilute gas in the corona 
may then become much higher than the kinetic energy per charged particle of the 
more dense fluid in the photosphere. In terms of the connection between temperature 
and kinetic energy, the temperature of the dilute gas in corona will be much higher 
than the temperature of the more dense fluid photosphere. 

If the kinetic energy of the circulating currents in that part of flux tubes floating 
in the corona becomes sufficiently high, then the flux tubes can explode violently into 
a solar flare which may be accompanied by a coronal mass ejection. 
The loss of magnetic energy during the flux tube explosion is rapidly converted into 
charged particle kinetic energy. The relativistic high energy products of the explosion 
yield both nuclear and elementary particle interactions. These processes are discussed 
in Sec.\ref{flare}. For magnetic flux tubes of smaller diameter which do not explode 
into a flare and/or a coronal mass ejection, one may still have low energy nuclear 
reactions that occur in a roughly steady state by continual conversion of magnetic 
field energy into charged particle energy. Such processes can account for the fact 
that the solar corona remains continually much hotter than the photosphere.
Steady state low energy nuclear processes are discussed in Sec.\ref{lenr}. In the 
concluding Sec.\ref{conc} we further discuss the notion that not all nuclear 
processes necessarily take place near within the solar core.

\section{\label{flare} Solar Flares}

\begin{figure}[bp]
\scalebox {0.6}{\includegraphics{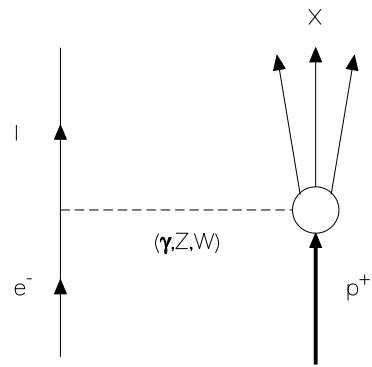}}
\caption{Boson exchange diagrams for electron-proton scattering 
into a lepton plus ``anything'' $\{\ e^- +p^+ \to l+X\ \}$ include photon 
$\gamma $ and $Z$ exchange wherein the final lepton is an electron,  
as well as charged $W^-$ exchange wherein the final state lepton is a neutrino. 
On an energy scale of $\sim 300\ {\rm GeV}$, all of these exchange processes have 
amplitudes of similar orders of magnitude.}
\label{fig2}
\end{figure}

The magnetic flux through a cylindrical tube of inner cross sectional area  
\begin{math} \Delta S \end{math} and mean magnetic field \begin{math} B \end{math}, 
is 
\begin{equation}
\Delta \Phi =B\Delta S.
\label{flare1}
\end{equation}
If a tube explodes in a time period \begin{math} \Delta t \end{math}, 
then the resulting loss of magnetic flux yields a mean Faraday law 
accelerator voltage around the tornado walls as given by  
\begin{equation}
\overline{\rm V}=\frac{\Delta \Phi}{\Delta t},
\label{flare2}
\end{equation}
i.e. 
\begin{equation}
e\overline{\rm V}=ecB\left(\frac{\Delta S}{\Lambda }\right) 
\ \ \ {\rm wherein}\ \ \ \Lambda=c\Delta t.
\label{flare3}
\end{equation}
A useful identity for numerical estimates is 
\begin{equation}
ecB\equiv 29.9792458
\left[\frac{\rm GeV}{\rm kilometer}\right]
\left[\frac{B}{\rm kiloGauss}\right].
\label{flare4}
\end{equation}
For a coronal mass ejection exploding coil with a Faraday flux loss 
time \begin{math} \Delta t \sim 10^2\ {\rm second}  \end{math} and 
with substantial sun spots at the ends of the magnetic flux coil, one 
may estimate\cite{Dikpati:2002,Lozitska:1994,Benz:2008}    
\begin{eqnarray}
\Delta S\approx \pi R^2 ,
\nonumber \\ 
R\sim 10^4\  {\rm kilometer}, 
\nonumber \\ 
B\sim 1\ {\rm kiloGauss},
\nonumber \\ 
\Lambda \sim 3\times 10^7\ {\rm kilometer}, 
\nonumber \\  
e\overline{\rm V}\sim 300\ {\rm GeV}.
\label{flare5}
\end{eqnarray}

The uncharged walls of the circulating vortex are represented roughly by  
an electron beam circulating in one direction and proton beam circulating 
in the other direction. These two colliding beams are hit with a flare or 
coronal mass ejecting Faraday law voltage pulse, as in Eq.(\ref{flare5}), 
setting an electron proton collision energy scale of 
\begin{math} E\sim 300\ {\rm GeV}  \end{math}. At such a high energy 
scale, electron-proton scattering\cite{Roberts:1990} is ruled by electro-weak 
exchange interactions all of the same order of magnitude in probability. 
Shown in FIG.\ref{fig2} is the electro-weak boson exchange Feynman diagram 
for electron-proton scattering  
\begin{equation}
e^- + p^+ \to l+X.
\label{flare6}
\end{equation}
The final state lepton is an electron for the case of photon 
\begin{math} \gamma \end{math} or \begin{math} Z \end{math} 
exchange and the final state lepton is neutrino for the case  
of \begin{math} W^- \end{math} exchange.

A solar flare or coronal mass ejecting event is thereby accompanied 
by an increased emission of solar neutrinos over a broad energy scale 
as well as relativistic protons\cite{Bostanjyan:2007}, 
neutrons\cite{Hurford:2006,Murphy:1999,Hua:2002,Murphy:2007, Watanabe:2003} 
and electrons\cite{Kahler:2007}. The full 
plethora\cite {Ryan:2006,Li:2007} of final \begin{math} X \end{math} 
states including electron, muon and pion particle anti-particle pairs 
should also be present in such events. The conversion of magnetic field 
energy into relativistic particle kinetic energy via the Faraday law voltage 
pulse is collective in that the magnetic flux in the core of the vortex depends 
on the rotational currents of {\em all} of the initial protons and electrons.

\section{\label{lenr} Low Energy Nuclear Reactions}
 
Even without spectacular solar flare explosions ejecting mass through the 
solar corona, there exists (within flux tubes) collective magnetic 
energy which allows for a significant occurrence of low energy nuclear reactions 
at many different locations in and around the sun. 
In particular, let us consider the inverse beta decay reaction 
\begin{equation}
W_{\rm magnetic}+e^- + p^+ \to \nu_e + n
\label{lenr1}
\end{equation}  
wherein the final state lepton is a neutrino, the final state 
``\begin{math} X \end{math}'' is a neutron \begin{math} n \end{math} 
and \begin{math} W_{\rm magnetic} \end{math} is magnetic field energy 
fed into the reaction.  

In a steady state flux tube (which does {\em not} explode) entering 
into the solar corona from one sunspot and exiting out of the solar 
corona through another sunspot, there is a substantial amount of 
stored magnetic energy. If there 
is a small change \begin{math} \delta I \end{math} in the current going 
around the vortex circumference, then the small change in the magnetic 
field energy 
\begin{math} \delta {\cal E}  \end{math} obeys 
\begin{equation}
\delta {\cal E}=\Phi \delta I. 
\label{lenr2}
\end{equation}
If \begin{math} L  \end{math} denotes the length of the vortex circumference of the   
magnetic flux tube, then the change in current due to the weak interaction reaction 
Eq.(\ref{lenr1}) is given by 
\begin{equation}
\delta I=-\frac{ev}{L}\ ,
\label{lenr3}
\end{equation}
wherein \begin{math} v \end{math} is the relative velocity component 
(tangent to the circumference) between the proton and electron. 
Putting \begin{math} \Phi = B\Delta S  \end{math} and 
\begin{math} \delta {\cal E}=-W_{\rm magnetic} \end{math} yields 
\begin{equation}
W_{\rm magnetic}=ec\left(\frac{\Phi }{L}\right)\frac{v}{c}=
ecB\left(\frac{\Delta S}{L}\right)\frac{v}{c}\ .
\label{lenr4}
\end{equation}
The product \begin{math} (ecB)  \end{math} is given in Eq.(\ref{flare4}). 
For the case of a cylindrical flux tube, 
\begin{equation}
\frac{\Delta S}{L}=\left(\frac{\pi R^2}{2\pi R}\right)=\frac{R}{2}\ .
\label{lenr5}
\end{equation}
yielding 
\begin{equation}
W_{\rm magnetic}\approx 15{\rm \ GeV} 
\left[\frac{R}{\rm kilometer}\right]
\left[\frac{B}{\rm kiloGauss}\right]
\frac{v}{c}\ .
\label{lenr6}
\end{equation}
Employing the estimates 
\begin{eqnarray} 
R\sim 10^2\  {\rm kilometer}, 
\nonumber \\ 
B\sim 1\ {\rm kiloGauss},
\nonumber \\ 
\frac{v}{c}\sim 10^{-2}, 
\nonumber \\ 
W_{\rm magnetic}\sim 15\ {\rm GeV}.
\label{lenr7}
\end{eqnarray}
On the energy scale \begin{math} W_{\rm magnetic} \ll 300\ {\rm GeV} \end{math} of 
Eq.(\ref{lenr7}), the weak interaction \begin{math} p^+ e^-   \end{math}  
processes Eq.(\ref{lenr1}) that produce neutrons proceed  more slowly than the purely 
electromagnetic \begin{math} p^+ e^- \end{math} processes. Nevertheless one finds 
appreciable neutron production in the solar corona. The production of neutrons among 
the protons allows for the creation of nuclei with higher mass numbers via neutron-capture 
nuclear reactions and subsequent beta decays.

\section{\label{conc} Conclusion}

Magnetic flux tubes arise\cite{Dikpati:2002,Lozitska:1994} out of
the turbulent magneto-fluid mechanics of a solar fluid plasma with high electrical 
conductivity. Turbulent magneto-fluid flows yield a full spectrum of magnetic 
field values\cite{Benz:2008} expected to vary randomly over many different 
length scales. The estimates of the magnetic field 
discussed in this work are merely order of magnitude. They are based on observations 
of the magnetic flux tubes entering into and exiting out of sunspots and also into and 
out of smaller crevices and holes that are commonly observed on the sun's optical surface.

For many years, the source of relativistic particle 
fluxes\cite{Yousef:2005,Roussev:2004} often observed to emanate from the solar corona 
has been theoretically obscure. Our explanation  for these fluxes is simply derived from 
Faraday's law 
\begin{equation}
-\frac{\partial {\bf B}}{\partial t}=curl{\bf E}
\label{conc1} 
\end{equation} 
as it appears in well understood transformer and inductor circuits. Circulating currents
around the walls of a flux tube can transfer energy into some parts of the magnetic field 
configuration while removing equal amounts of energy from other distant parts of the 
magnetic field configuration. This transformer action is very well understood.

The resulting energy balance allows a large number of low energy charged particles to 
collectively transfer  their kinetic energy to a significantly smaller number of charged 
particles whose energy per particle then becomes very high. When the charged particle 
energy of low density solar corona particles is made sufficiently high, reactions of 
the form in Eqs.(\ref{flare6}) and (\ref{lenr1}) clearly can take place, 
leading to neutron production. Once neutrons are created and added to the 
electron-proton plasma, a variety of nuclear synthesis reactions 
become possible\cite{Fowler:1965}.  If that is the case, then Coulomb barrier-penetrating 
fusion reactions in the sun's core are not necessarily the sun's {\em only} significant 
source of solar nuclear energy. Relativistic particle fluxes have been clearly observed 
emanating from regions located well above the sun's surface and very far away 
from the solar core. Finally, it has not escaped our notice that the energetic particle 
production via the collective mechanisms discussed in this work may shed some light on the 
origin of the anomalous short-lived isotopes observed on other astronomical 
objects\cite{Cowley:2004}.

\end{document}